# Page Layout Analysis of Text-heavy Historical Documents: a Comparison of Textual and Visual Approaches

Sven Najem-Meyer[1,*], Matteo Romanello[2]

[1]EPFL, Lausanne, Switzerland.
[2]UNIL, Lausanne, Switzerland

## Abstract

Page layout analysis is a fundamental step in document processing which enables to segment a page into regions of interest. With highly complex layouts and mixed scripts, scholarly commentaries are text-heavy documents which remain challenging for state-of-the-art models. Their layout considerably varies across editions and their most important regions are mainly defined by semantic rather than graphical characteristics such as position or appearance. This setting calls for a comparison between textual, visual and hybrid approaches. We therefore assess the performances of two transformers (LayoutLMv3 and RoBERTa) and an objection-detection network (YOLOv5). If results show a clear advantage in favor of the latter, we also list several caveats to this finding. In addition to our experiments, we release a dataset of ca. 300 annotated pages sampled from 19th century commentaries.

*Corresponding author.
sven.najem-meyer@epfl.ch (S. Najem-Meyer); matteo.romanello@unil.ch (M. Romanello)
0000-0002-3661-4579 (S. Najem-Meyer); 0000-0002-7406-6286 (M. Romanello)

# 1. Introduction

## 1.1. Page layout analysis

Automatically transcribing a page by means of optical character recognition (OCR) often results in losing crucial information about its layout. This loss can be critical for further analyses which typically require accessory regions such as running headers and footnotes to be separated from the main text. Similarly, capturing information about page layout is of key importance for the automatic or semi-automatic markup of digitized documents, as textual information contained in each page region can be automatically marked up, provided that a mapping is established between region types and markup elements.

To tackle this issue, we focus on Page Layout Analysis[1], which aims at segmenting a page into homogeneous regions and at classifying those regions according to their contents [1, 2]. Region contents can be of both textual and visual nature, and the two modalities can be leveraged in a separate or combined fashion. Purely textual approaches construe layout analysis as a natural language processing (NLP) problem. They aim at delimiting and at labeling the sequence of text composing a region. Visual approaches, on the other hand, seize the task as a computer vision problem and aim at detecting and classifying image regions. Finally, hybrid approaches leverage both modalities to detect and classify image regions and their corresponding text sequence.

Visual approaches are often considered the standard way to go. This trend is probably encouraged by the recent progress of pre-trained convolutional neural networks (CNN) and by their ability to deal with non-textual regions. These approaches unsurprisingly show their best performances in distinguishing regions with highly contrasting graphical attributes, such as tables, illustrations and drop capitals. Yet, regions are often characterised by semantic rather than graphical features. In this case, it makes sense to opt for textual or hybrid approaches. If purely textual approaches prove their usefulness when a page's image is not aligned to its text or not available at all (e.g. [3]), they end up discarding relevant information when it is. Hybrid models hence make use of images, text and their corresponding coordinates. Notice that this can be done either by enhancing an image-based model with text embeddings (e.g. by addition or concatenation) or by providing a textual model with text-coordinates in parallel to a visual backbone. How these three approaches are best suited to analysing text-heavy documents remains to be addressed.

## 1.2. Background: the case of classical commentaries

In this paper, we focus on historical classical commentaries. We place them in the broader category of text-heavy documents as they mostly contain text, as opposed to more visual documents like illuminated manuscripts. The research project Ajax Multi-Commentary[2] serves as the context for this work. It aims to create an automated pipeline to convert digitized commentaries into a body of structured information to aid in their comparative analysis. Within this pipeline, page layout analysis plays a crucial role as it can enable the (semi-)automatic markup of information contained within commentaries.

[1]We use *Page Layout Analysis* rather than the more generic *Document Layout Analysis* because the latter includes recognizing regions that span multiple pages, which is beyond the scope of this study.
[2]https://github.com/AjaxMultiCommentary

Together with critical editions and translations, commentaries are one of the main genres of publication in literary and textual scholarship. Providing in-depth analyses often side by side with a critical edition of the chosen text, commentaries can have very sophisticated layouts with considerable variations between editions (Fig. 1). A common layout type has the commentary section as a single or double column footnote section positioned below the primary text or its translation. In other layout types, however, commentary sections span over the entire page. Conversely, regions with similar placement and appearance can have different functions. Besides the complexity of their layout, commentaries also feature a specific prose style, clearly recognizable by its intertwining of multiple scripts and its pervasive use of abbreviations. Comments generally follow a determined pattern such as *line number - commented word or excerpt - comment*, for instance "1 *(line)*. Ἀεὶ μέν...καὶ νῦν *(Excerpt)*: cp. Tr. 689-691. The passage in Aesch. Ag. 587-598 is scarcely a true parallel [...] *(comment)*".

As our project's pipeline starts with commentary images and ends with text mining, we value page layout analysis as a crucial step in which primary text, margin notes, line numbers and commentary sections ought to be precisely segmented. This task remains a challenging one given the characteristics listed above. If information about layout is mainly conveyed by semantic rather than by graphical clues, visual features are not irrelevant. Commentary regions are generally written in a smaller font and are often punctuated by bold line numbers anchors. Besides, for Greek commentaries, the script lends to a good visual feature to differentiate between the primary text and its translation.

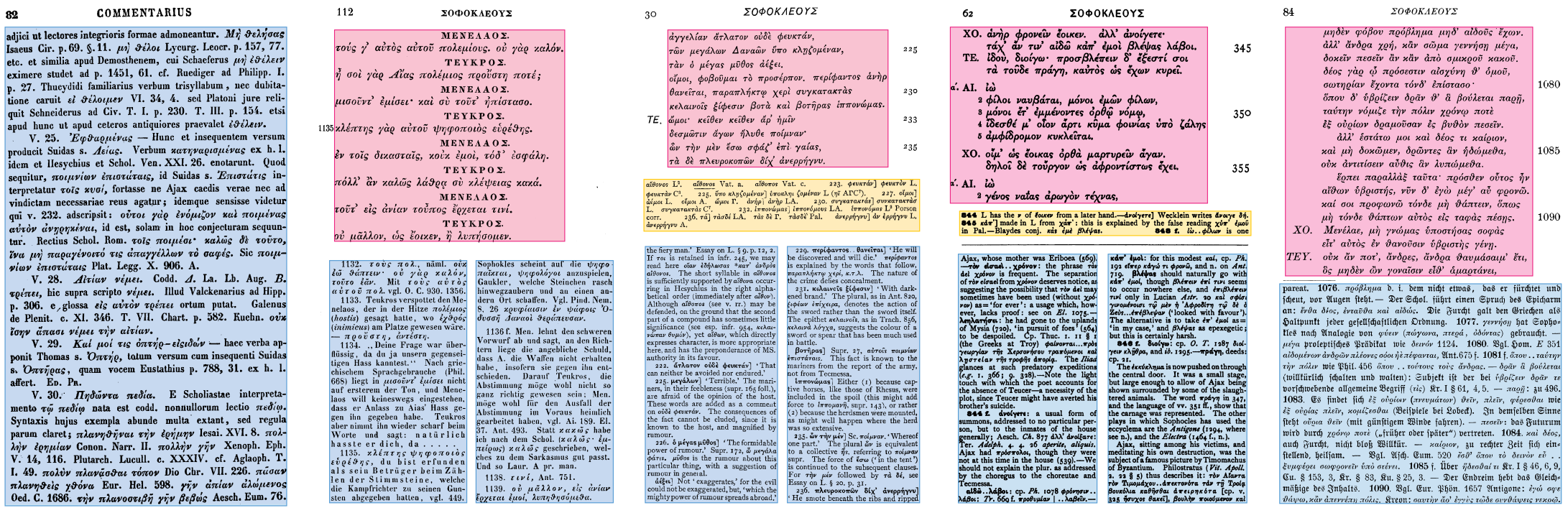

**Figure 1:** Example pages from 19th century commentaries on Sophocles' *Ajax*. The commentaries are by (from left to right): Lobeck (1835), Schneidewin (1853), Campbell (1881), Jebb (1896) and Wecklein (1894). Commentary sections are highlighted in blue, primary texts in pink and critical apparatus in yellow.

## 1.3. Goals

In this challenging and mostly uncharted setting, our primary goal is to assess the performances of textual, visual and hybrid approaches. For each of these approaches, we ask the following questions:

- **RQ1**: How well do state-of-the art models perform over commentaries of works written in different scripts (e.g. Latin or Polytonic Greek), belonging to different literary genres and having different layout types? Which of the textual, visual and hybrid approaches is best suited for the task?
- **RQ2**: What is the impact of the quantity of training data?
- **RQ3**: How well do models generalize on layout type they have not seen during training?

For textual and hybrid approaches only, we address two additional questions:

- **RQ4**: How do hybrid models perform on languages they have not been pre-trained on?
- **RQ5**: To which extent do textual and visual features separately account for the model's decision?

## 2. Related works

Generic approaches to page layout analysis have known considerable progress in recent years. Overtaking CNNs, image transformers such as DiT [4] or LayoutLMv3 [5] can be used for several visual or multi-modal document analysis tasks. However, on the contrary to newspapers, magazines and scientific press, commentaries remain barely explored as far as layout analysis is concerned. We therefore compared studies on historical documents, as they share the many similarities with commentaries.

Simistra et al. [6] report the performances of several pixel classification algorithms for the Competition on Layout Analysis for Challenging Medieval Manuscripts at ICDAR 2017. The tasks include region detection for text, comments and images. Results show a net advantage in favor of convolutional neural networks (CNN), with intersection over union (IoU) scores ranging up to .90 for comments. It must be signaled, however, that comments consistently take the form of of marginal glosses and thereby possess very distinctive graphical features.

Mehri et al. [7] also report performances of pixel-based approaches for the Competition on Historical Book Analysis at ICDAR 2019. The competition is based on two challenges: distinguishing between text and images and classifying various text fonts. The best models used fully CNNs and reach scores close to perfection in the first challenge (.99 F-score and above), but the binary classification is relatively easy for this type of pre-trained networks. Results to the second challenges, though slightly lower, are particularly interesting to our research. They show that CNN can leverage fine-grained information regarding the font style (e.g. bold, italicized, etc.), which may avail in our case.

Finally, Yang et al. [8] proposed a multimodal CNN to extract semantic structure from documents. The principle is to build a text embedding map which is accessible by the last layer of the model. Building on this idea, Barman et al. [9] reports notable improvement when using textual features to segment historical newspapers. If the text-only models yield the lowest scores, combining text and images consistently outperform image-only features by 3% mIoU in average.

## 3. Datasets

While ground truth datasets already exist for layout analysis of historical documents such as manuscripts and early printed books [10, 11], newspapers [12] and even for the semantic segmen-tation of geographical maps [13], no such dataset existed for scholarly commentaries or critical editions. We contribute to filling this gap by creating and releasing GT4HistCommentLayout, a dataset of page layout analysis annotations on 19th century commentaries to Ancient Greek and Latin works, written in English, French and German[3]. This new dataset complements GT4HistComment [14], which provides OCR ground-truth data for the same type of historical documents.

## 3.1. Layout annotation

To perform layout annotation we devise a content-based region taxonomy geared towards commentaries and critical editions (Fig. 2). It consists of 18 fine-grained classes, which are mapped to 8 coarse-grained classes in order to reduce the class number and class sparsity. Mapping is achieved by grouping region types with similar visual characteristics (e.g. numbers). The list of classes defined by our taxonomy is given in Table 1. For the experiments reported below, we exclusively consider coarse-grained classes.

**Table 1**
Complete list of fine- and coarse-grained page region classes used for layout annotation, with their corresponding mapping to SegmOnto's controlled vocabulary.

| Fine | Coarse | SegmOnto Type:Subtype |
|---|---|---|
| commentary | commentary | MainZone:commentary |
| critical apparatus | critical apparatus | MarginTextZone:criticalApparatus |
| footnotes | footnotes | MarginTextZone:footnotes |
| page number | number | NumberingZone:pageNumber |
| text number | number | NumberingZone:textNumber |
| bibliography | others | MainZone:bibliography |
| handwritten marginalia | others | MarginTextZone:handwrittenNote |
| index | others | MainZone:index |
| others | others | CustomZone |
| printed marginalia | others | MarginTextZone:printedNote |
| table of contents | others | MainZone:ToC |
| title | others | TitlePageZone |
| translation | others | MainZone:translation |
| appendix | paratext | MainZone:appendix |
| introduction | paratext | MainZone:introduction |
| preface | paratext | MainZone:preface |
| primary text | primary text | MainZone:primaryText |
| running header | running header | RunningTitleZone |

[3]https://doi.org/10.5281/zenodo.7271729

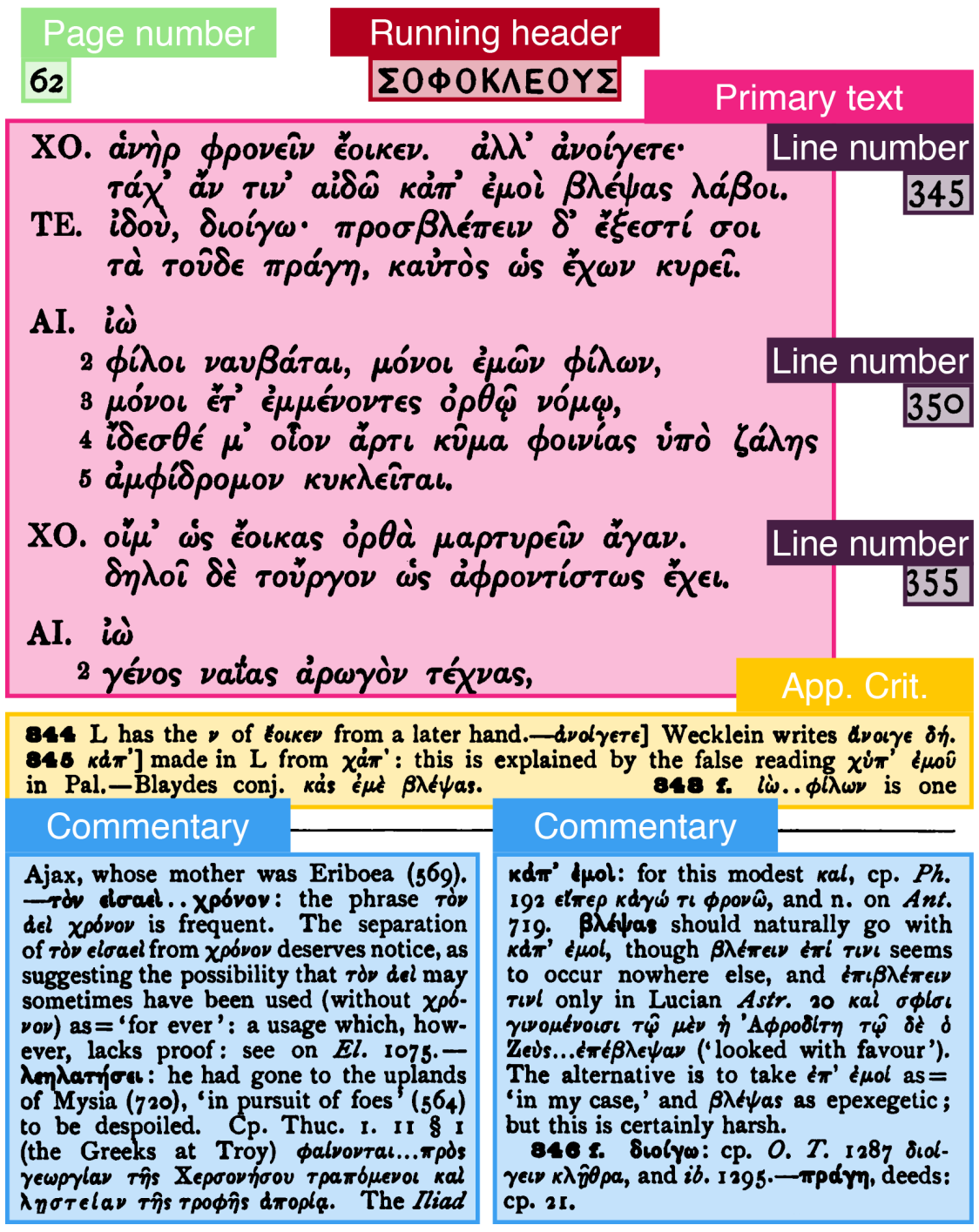

Figure 2: The main layout elements of a scholarly commentary page.

This taxonomy distinguishes between the original Greek text of the work being commented upon (*Primary text*), the commentary sections (*Commentary*), the commentator's translation of the commented text (*Translation*), the section containing information about manuscript readings and editorial conjectures (*Critical apparatus*), the paratextual elements — e.g. table of contents, appendices, indices, footnotes, introductory and prefatory materials — (*Paratext*) and finally page and line numbers (*Number*).

In order to make the published dataset as widely reusable as possible, we mapped our classes to the SegmOnto controlled vocabulary [15, 16]. The only difficulty we encountered in the mapping to SegmOnto concerned the *commentary* region class, as it can be mapped both to a `MainZone` or to a `MarginTextZone`, depending on the commentary at hand. In fact, in commentaries containing both primary text and commentary, commentary regions could be interpreted as *marginalia* to the commented text (i.e. a `MarginTextZone`); whereas in commentaries with no primary text, the commentary itself is undoubtedly the main region of the page (i.e. a `MainZone`). We address this issue by always considering *commentary* of type `MainZone`, based on the consideration that the area of the page such regions tend to occupy is roughly equal to the area of *primary text* or *translation* regions (when present).

Annotation was performed by three annotators by using the VGG Image Annotator (VIA) tool [17]. While each commentary was annotated by one person at a maximum, all annotations were revised by an expert in order to ensure consistency in the application of the annotation guidelines. Manually annotated page regions were automatically resized to fit exactly the minimal bounding rectangle around contained words.

### 3.2. Sampling and dataset composition

As a sampling strategy, we started with ca. 40 pages of annotation per commentary. We made sure that all page layout types (see Fig. 3 for selective examples) of any given commentary are included in the sample because page layout can vary quite substantially throughout a commentary depending on the section contents.

The data used for experiments consist of an *internal* and an *external dataset*. The *internal dataset* comprises of commentaries to Sophocles' *Ajax*, published from the beginning of the 19$^{\text{th}}$ century to date. Of these 12 commentaries, slightly less than a half are in the public domain, while the remaining are still under copyright. The *external dataset*, instead, consists of commentaries to both Latin and Greek classical works, sampled to include works both in prose and poetry. It contains an English commentary to Tacitus' *Annals* (Latin prose), a German commentary to book 6 of Vergil's *Aeneid* (Latin poetry), and a German commentary to book 7 of Thucydides' *History of the Peloponnesian War* (Ancient Greek prose). The specific purpose of this external dataset is to evaluate with which accuracy layout analysis models trained on data from one specific genre and literature (i.e. Greek poetry, in the case of the *Ajax*) can be applied to commentaries about works from a wider variety of literary genres (see RQ6).

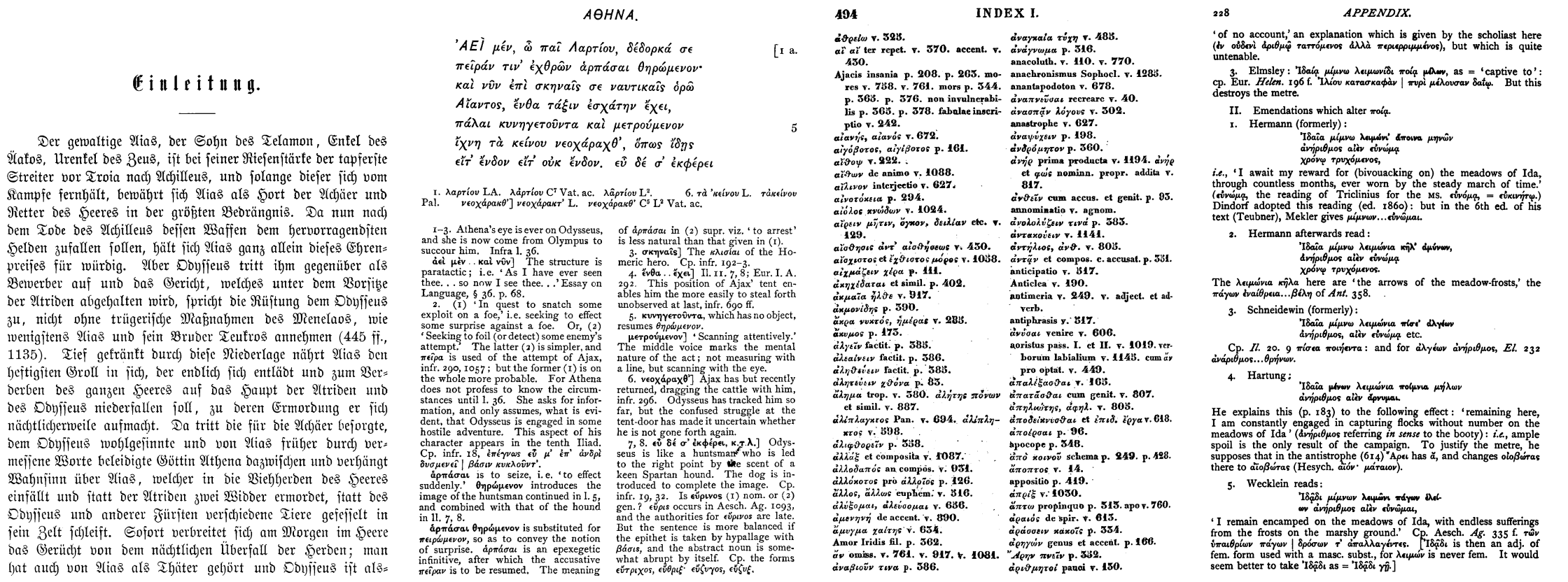

**Figure 3:** Overview of various layouts. From left to right: Introduction (from Wecklein), Commentary and primary text (from Campbell), index (from Lobeck) and appendix (from Jebb).

Given this important distinction, the ground-truth dataset we release contains the public domain portion of the internal dataset, as well as the entire external dataset (as it consists exclusively of out of copyright documents). Detailed statistics about these datasets can be found in Table 2.

## 4. Experimental setup

### 4.1. Models

**LayoutLMv3** For hybrid experiments, we use LayoutLMv3$_{BASE}$ [5], a transformer which uses text, text-coordinates and image as inputs. This choice is motivated by the need to have a state-of-the-art hybrid model easily comparable both with a textual approach (by pitting it against RoBERTa, infra) and a visual approach (by way token ablation). On the contrary to its predecessor, LayoutLMv3 does not rely on a pre-trained CNN for its visual backbone, but uses a multi-modal transformer instead. The authors claim superior results to concurrent systems such as DocFormer or StructuralLM. Pretests showed LayoutLMv3 to be slightly superior to LayoutLMv2 at the cost of a longer training time. As the model converged after 30 epochs, we fine-tune each model for a total of 40 epochs using recommended parameters and a maximum length of 512 tokens per instance. In the experiments below, we use LayoutLM for token classification, which opens three possible ways of labeling the data. The first method consists in annotating only the first word of a region. This method has the downside of creating highly imbalanced classes, with a vast majority of words marked with a zero-label and very few marked with their region's class. This method did not yield encouraging results in pre-tests and was therefore abandoned. The second method is inspired by the named entity recognition field and consists in labelling the first word of a region with `BEGIN-[RegionClass]` and the following with `INSIDE-[RegionClass]`. Besides doubling the number of classes, this method leads to the creation of very long entities and performed poorly in pre-tests. We therefore go for the third method, which consists in labeling all the words in a region with the regions label.

**Table 2**
Detailed statistics about the annotated data. For each annotated commentary we report the number of pages as well as the total number of regions per class.

| Commentary | Pages | AppCrit | Comm. | Footn. | Num. | Others | Parat. | Primary t. | Running h. |
|---|---|---|---|---|---|---|---|---|---|
| *Internal commentaries (public domain)* | | | | | | | | | |
| Lobeck 1835 | 61 | 0 | 20 | 13 | 227 | 32 | 61 | 6 | 67 |
| Campbell 1881 | 42 | 26 | 52 | 11 | 112 | 20 | 42 | 16 | 26 |
| Jebb 1896 | 43 | 25 | 50 | 8 | 87 | 55 | 43 | 11 | 18 |
| Schneidewin 1853 | 62 | 0 | 84 | 3 | 126 | 10 | 62 | 20 | 42 |
| Wecklein 1894 | 42 | 0 | 35 | 2 | 145 | 12 | 42 | 5 | 41 |
| **Total** | 250 | 51 | 241 | 37 | 697 | 129 | 250 | 58 | 194 |
| *Internal commentaries (under copyright)* | | | | | | | | | |
| Colonna 1975 | 40 | 28 | 0 | 10 | 164 | 12 | 40 | 12 | 26 |
| De Romilly 1976 | 41 | 28 | 33 | 4 | 140 | 18 | 41 | 8 | 30 |
| Ferrari 1974 | 40 | 0 | 57 | 8 | 111 | 15 | 40 | 9 | 29 |
| Garvie 1998 | 40 | 9 | 10 | 6 | 136 | 15 | 40 | 7 | 10 |
| Kamerbeek 1953 | 40 | 0 | 30 | 12 | 38 | 9 | 40 | 10 | 0 |
| Paduano 1982 | 40 | 0 | 22 | 0 | 139 | 20 | 40 | 9 | 15 |
| Untersteiner 1934 | 40 | 0 | 27 | 0 | 76 | 16 | 40 | 7 | 26 |
| **Total** | 281 | 65 | 179 | 40 | 804 | 105 | 281 | 62 | 136 |
| *External commentaries (public domain)* | | | | | | | | | |
| Classen & Steup 1889 | 41 | 0 | 44 | 0 | 74 | 3 | 19 | 22 | 37 |
| Norden 1903 | 40 | 10 | 16 | 2 | 107 | 18 | 6 | 9 | 38 |
| Furneaux 1896 | 40 | 30 | 60 | 8 | 140 | 44 | 5 | 31 | 37 |
| **Total** | 121 | 40 | 120 | 10 | 321 | 65 | 30 | 62 | 112 |
| **Grand total (public domain)** | 371 | 91 | 361 | 47 | 1018 | 194 | 280 | 120 | 306 |
| **Grand total (all)** | 652 | 156 | 540 | 87 | 1822 | 299 | 561 | 182 | 442 |

**RoBERTa** Provided the multilingualism of commentaries, it could have been relevant to use a multilingual transformer for text-only experiments. However, as LayoutLMv3 uses RoBERTa [18] to initialize its embeddings, we stick to the method used by its authors [5] and chose to train RoBERTa$_{BASE}$ for a fair comparison with the former. This bi-directional multi-head attention transformer was released as an improved version of BERT [19], being pre-trained on 160GB of uncompressed English text from Wikipedia, BooksCorpus, CC-News, OpenWebText and Stories [18]. Regarding training and labelling, we use the same settings as LayoutLM.

**YOLOv5** For visual experiments, we use YOLOv5[4,5], an object-detection model based on DarkNet. The choice of YOLO is mainly motivated by its encouraging results in historical document layout recognition [20]. In preliminary tests, the model performed best with an image resolution of 1280 and converged around epoch 250. Regarding the size of the model, the larger version (YOLOv5x) did not yield considerably better results despite a much longer training time. All experiments are therefore run on YOLOv5m with a resolution of 1280 for 300 epochs. In order to assess the amount of difficulty added by multiplying classes, we create two YOLO models:

- YOLO$_{Mono}$ is trained for single class object detection, which is enabled by labelling all regions identically.
- YOLO$_{Multi}$ is trained for multi-class object detection, using our dataset's coarse labels (see Section 3.1).

**YOLO$_{Mono}$+LayoutLM/RoBERTa** This model combines the two approaches, using YOLO$_{Mono}$ to detect regions and LayoutLM/RoBERTa to classify them. Words contained within predicted regions are fed to LayoutLM/RoBERTa. The majority class among words is then used to label the regions.

## 4.2. Implementation and training

We implement our experiments using HuggingFace `transformers`[6] and YOLO's API[7]. Training was performed on two NVIDIA GeForce GTX TITAN X GPUS, each with 12.2GO of memory. The code is made publicly available on GitHub[8].

## 4.3. Evaluation methods

As LayoutLM and RoBERTa are used for token classification, they should be evaluated on entity or word basis. However, in order to enable a meaningful comparison with YOLO, we group consecutive words with identical labels and build up a region from their bounding rectangle. Notice that LayoutLM and RoBERTa are severely disadvantaged by this evaluation procedure, as a single incorrectly labelled word among an actual region disrupts its unity. This problem is illustrated Figure 4 in the appendix and is not straightforward to mitigate without visual operations or carefully tailored rules. Indeed, homogenising long strands of tokens could result in the absorption of tiny regions like line numbers. We therefore evaluate the results without post-processing them and compute all mean average precision (mAP) scores at a 0.5 IoU threshold[9]. It is worth noting that the obtained scores are approximately .10 mAP points

[4]https://github.com/ultralytics/yolov5

[5]Despite multiple attempts, we couldn't get Kraken's (https://github.com/mittagessen/kraken) segmentation training to work on our infrastructures and therefore removed it from our experimental procedure.

[6]https://github.com/huggingface/transformers

[7]https://github.com/ultralytics/yolov5

[8]https://github.com/AjaxMultiCommentary/ajmc/tree/main/ajmc/olr

[9]We used the Python package `mean-average-precision`, https://github.com/bes-dev/mean_average_precision, version 2021.4.26.0

below the scores produced by YOLO's built-in evaluation tool, a discrepancy already mentioned by [11][10].

## 5. Experiments

We divide our experiments according to our research questions and list them in Table 3. As using only textual features is consistently reported to yield lower results [9, 8, 5], we test this approach in a single sub-experiment to the hybrid series. This allows us to simplify our experimental design and to spare computing power while still being able to measure the benefits of adding image and coordinates. Results are presented in Table 4, and sample predictions are shown in Figure 4.

**Table 3**
Experimental design.

| id | name | RQ | Train data | Test data | Languages |
|---|---|---|---|---|---|
| 0A | Jebb - Base | RQ1 | Jebb | Jebb | en, gr |
| 0B | Kamerbeek - Base | RQ1 | Kamerbeek | Kamerbeek | en, gr |
| 1A | Jebb - Half trainset | RQ2 | Jebb | Jebb | en, gr |
| 1C | Jebb - Token ablation | RQ5 | Jebb | Jebb | - |
| 1D | Jebb, Kamerbeck - base | RQ3 | Jebb, Kamerbeek | Jebb | en, gr |
| 1E | Jebb - Text only | RQ1&5 | Jebb | Jebb | en, gr |
| 2A | Campbell, Jebb - Transfer | RQ3 | Campbell | Jebb | en, gr |
| 2B | Kamerbeek, Jebb - Transfer | RQ3 | Kamerbeek | Jebb | en, gr |
| 2C | Garvie, Jebb - Transfer | RQ3 | Garvie | Jebb | en, gr |
| 3A | Paduano - Base | RQ4 | Paduano | Paduano | it, gr |
| 3B | Wecklein - Base | RQ4 | Wecklein | Wecklein | de, gr |
| 4A | Omnibus (internal) | RQ1 | All (internal) | All (internal) | en, de, it, lat, gr |
| 4B | Omnibus (external) | RQ1 | All (external) | All (external) | en, de, lat |
| 4C | Omnibus - Transfer | RQ3 | All (internal) | All (external) | en, de, lat |

**RQ1: Which of the textual, visual and hybrid approaches performs best?** As described in Section 1.3, our primary goal is to assess the performances of state-of-the-art models on classical commentaries and to investigate which of the three named approaches is the most appropriate for this kind of data.

**Experimental design.** As LayoutLM is pre-trained on English data, we first test its performances on two English commentaries: Jebb's (baseline, experiment 0A) and Kamerbeek's (0B). Besides its scholarly resonance, we chose Jebb's commentary as a baseline because it contains regions of all coarse classes. As for Kamerbeek's commentary, it presents an utterly different layout in which the commentary sections span over an entire page.

[10]See also https://github.com/bes-dev/mean_average_precision/issues/1

Additionally, we also train and test our models on a diverse set of commentaries on Sophocles' *Ajax* (experiment 4A) and on other Greek and Latin prose and poetry works (4B). We then test visual approaches by running the same experiments with YOLO$_{Multi}$. Finally, we test textual approaches with RoBERTa using the same data as 0A (experiment 1E).

**Results and Discussion.** Results show a net advantage in favor of YOLO$_{Multi}$, which overtakes LayoutLM by an average of .27 points over experiments 0A, 0B, 4A and 4B. Interestingly and on the contrary to LayoutLM, YOLO$_{Multi}$ completely misses footnotes in Jebb (N=8) and systematically incorporates them within the main paratext region. As for RoBERTa, its poor results are inline with previously mentioned studies showing the inferiority of text-only approaches. This first series of experiments shows that image-based approaches can perform well even on region with few distinctive graphical features if they have seen similar layouts in training.

**RQ2: What is the impact of training set's size?** To address this question, we copy the setting of experiment 0A, only changing the size of the training set by sampling half of it randomly (experiment 1A).

**Results and Discussion.** If both YOLO$_{Multi}$ and LayoutLM show a performance drop in comparison with 0A, it is worth noting that depriving the former of half its training data only leads to a .05 decrease in mAP. The latter's case is more concerning and deserves a more thorough inquiry. First, the model did not seem to be penalised by the number of epochs, as its maximum score is already attained at epoch 33/50. Secondly, the difference in mAP does not reflect the difference in word-based F1-score, which only decreases of .10. In-depth analyses revealed predictions to be much more scattered, which drastically hampers homogeneous region building and accounts for the plunge of mAP scores. The takeaway of this experiment is that 15 to 20 pages of ground-truth data already opens the way to satisfactory results, whereas doubling this amount only accounts for an improve of .05 mAP.

**RQ3: How well do models generalize on layout types they have not encountered during training?** We address the question of generalization in three ways. We first train a model on two commentaries and evaluate it on Jebb (baseline) to see whether mixing layout types can confuse the model. For this sub-experiment (1D) we use the commentaries by Jebb and Kamerbeek, two English commentaries with different layouts which we already have individual baselines for (cf. 0A and 0B). We then train three models on three English commentaries and evaluate them, again, on Jebb. We choose one commentary with an almost identical layout (Campbell, experiment 2A), and two commentaries with a completely different layout, Kamerbeek (2B) and Garvie (2C). In these two items, commentary sections cover the main zone of the page. Finally, we train a model on all internal commentaries and test it on external commentaries (4C).

**Results and discussion.** First, it seems that mixing two types of layout in training did not confuse the model. On the contrary, YOLO$_{Multi}$ shows a .15 increase in mAP between 0A and 1D. This improvement is probably due to the quantity of available data, as regions such as running headers, paratext, numbers and footnotes see their number of training instances doubled and their scores consistently improved. Interestingly, this correlation is not present for commentary sections. Its AP remains at .90 despite a rise in $N_t$ from 40 to 66. This plateau can be explained by the important change in the region's morphology between Kamerbeek and Jebb. More generally, this result suggests that using a single model with more data yields better results than individual models.

For experiments 2A, 2B and 2C, we generally observe a net decrease of performance when compared to the baseline. This being said, results are still better when generalizing to a similar layout type. Experiment 2A is therefore above 2B and 2C for both LayoutLM and YOLO$_{Multi}$. These results also hint at the fact that LayoutLM only seems to gain little information from the textual channel, a trend to be confirmed below. Performances also decrease in experiment 4C, despite the broadness of the training set. This result shows compelling evidence about the model's struggles with completely unseen data. Indeed, if many of the layout types are present in the training, one must not understate the importance of other image features like the quality of the scan, the binarization threshold and so forth. To circumvent this problem, it is maybe sufficient to add very few images from the target data in training or fine-tuning. We keep this hypothesis to be tested in future works.

**RQ4: How do hybrid models perform on languages they have not been pre-trained on?** We then measure the impact of the text's language by training two models on an Italian (Paduano) and a German commentary (Wecklein) respectively (experiments 3A and 3B).

**Results and discussion.** As it appears, LayoutLM does not seem to be impacted by the commentary's main language. If results on German data fail to equate those of Jebb, Italian data gets the best results for a single commentary overall. These results can be explained by the domain-specific prose style of commentary writing. As mentioned in Section 1.2, the text often patches Greek scripts, abbreviations, rare words and proper nouns together. To circumvent this unusual distribution, LayoutLM's tokenizer has to chunk words into extremely tiny pieces to match them to its vocabulary. It is therefore very frequent to see words fed to the model as sequences of single-character embeddings. This setting lessens the model capacity to rely on the knowledge acquired during pre-training and hence degrades its overall performances.

**RQ5: To which extent do textual and visual features separately account for the LayoutLM's decision?** To measure this last statement more precisely, we run LayoutLM in token ablation mode (1C), feeding the model only with null tokens, thereby constraining its weights to rely solely on coordinates and images.

**Results and discussion.** RoBERTa's poor results (1E) already indicate that LayoutLM is highly dependant on coordinates and images. Experiment 1C confirms this intuition and contributes to explaining the model's indifference towards language. As a matter of fact, blanking textual inputs only diminishes the models performances by .01 mAP. In some regions with consistent positioning, textual inputs are even worsening the results: this is the case with

**Table 4**

General Results table, where bold numbers are applied to the highest score in a single experiments. $N_t$ and $N_e$ indicate the counts of instances in train and evaluation set respectively. Dashes stand for na-values.

| | | All | App. crit. | | | Commentary | | | Footnote | | | Numbers | | | Others | | | Paratext | | | Primary text | | | Running h. | | |
|---|---|---|---|---|---|---|---|---|---|---|---|---|---|---|---|---|---|---|---|---|---|---|---|---|---|---|
| Exp | Model | mAP | AP | $N_t$ | $N_e$ | AP | $N_t$ | $N_e$ | AP | $N_t$ | $N_e$ | AP | $N_t$ | $N_e$ | AP | $N_t$ | $N_e$ | AP | $N_t$ | $N_e$ | AP | $N_t$ | $N_e$ | AP | $N_t$ | $N_e$ |
| 0A | LLM | .38 | .12 | 20 | 5 | .51 | 40 | 10 | **.50** | 6 | 2 | .33 | 63 | 24 | .32 | 29 | 14 | .20 | 7 | 4 | .34 | 14 | 4 | .76 | 29 | 11 |
| | $Y_{Mono}$ | **.81** | - | - | - | - | - | - | - | - | - | - | - | - | - | - | - | - | - | - | - | - | - | - | - | - |
| | Y+LLM | .45 | .45 | 20 | 5 | .70 | 40 | 10 | .00 | 6 | 2 | .67 | 63 | 24 | .19 | 29 | 14 | .40 | 7 | 4 | .50 | 14 | 4 | .70 | 29 | 11 |
| | $Y_{Multi}$ | .69 | **.60** | 20 | 5 | **.90** | 40 | 10 | .00 | 6 | 2 | **.81** | 63 | 24 | **.62** | 29 | 14 | **.95** | 7 | 4 | **.75** | 14 | 4 | **.89** | 29 | 11 |
| 0B | LLM | .22 | - | 0 | 0 | .05 | 26 | 4 | .12 | 10 | 2 | .50 | 32 | 6 | **.00** | 6 | 2 | .36 | 7 | 3 | - | 0 | 0 | .71 | 31 | 7 |
| | $Y_{Mono}$ | **.93** | - | - | - | - | - | - | - | - | - | - | - | - | - | - | - | - | - | - | - | - | - | - | - | - |
| | Y+LLM | .21 | - | 0 | 0 | .00 | 26 | 4 | .00 | 10 | 2 | .17 | 32 | 6 | **.00** | 6 | 2 | .61 | 7 | 3 | - | 0 | 0 | .90 | 31 | 7 |
| | $Y_{Multi}$ | .51 | - | 0 | 0 | **1.00** | 26 | 4 | **.25** | 10 | 2 | **1.00** | 32 | 6 | **.00** | 6 | 2 | **.83** | 7 | 3 | - | 0 | 0 | **1.00** | 31 | 7 |
| 1A | LLM | .14 | .04 | 20 | 5 | .08 | 40 | 10 | .03 | 6 | 2 | .08 | 63 | 24 | .01 | 29 | 14 | .08 | 7 | 4 | .02 | 14 | 4 | .76 | 29 | 11 |
| | $Y_{Mono}$ | **.74** | - | - | - | - | - | - | - | - | - | - | - | - | - | - | - | - | - | - | - | - | - | - | - | - |
| | Y+LLM | .35 | .27 | 20 | 5 | .40 | 40 | 10 | .00 | 6 | 2 | .70 | 63 | 24 | .08 | 29 | 14 | .44 | 7 | 4 | .17 | 14 | 4 | .76 | 29 | 11 |
| | $Y_{Multi}$ | .64 | **.60** | 20 | 5 | **.80** | 40 | 10 | **.10** | 6 | 2 | **.94** | 63 | 24 | **.45** | 29 | 14 | **.83** | 7 | 4 | **.42** | 14 | 4 | **1.00** | 29 | 11 |
| 1C | LLM | .37 | **.60** | 20 | 5 | **.80** | 40 | 10 | **1.00** | 6 | 2 | .01 | 63 | 24 | **.22** | 29 | 14 | .21 | 7 | 4 | .10 | 14 | 4 | **.05** | 29 | 11 |
| | Y+LLM | **.43** | .40 | 20 | 5 | .78 | 40 | 10 | **1.00** | 6 | 2 | **.22** | 63 | 24 | .19 | 29 | 14 | **.63** | 7 | 4 | **.17** | 14 | 4 | .03 | 29 | 11 |
| 1D | LLM | .26 | .15 | 20 | 5 | .34 | 66 | 10 | .00 | 16 | 2 | .35 | 95 | 24 | .22 | 35 | 14 | .07 | 14 | 4 | .11 | 14 | 4 | .85 | 60 | 11 |
| | $Y_{Mono}$ | .83 | - | - | - | - | - | - | - | - | - | - | - | - | - | - | - | - | - | - | - | - | - | - | - | - |
| | Y+LLM | .43 | .40 | 20 | 5 | .19 | 66 | 10 | .50 | 16 | 2 | .58 | 95 | 24 | .27 | 35 | 14 | .32 | 14 | 4 | .34 | 14 | 4 | .85 | 60 | 11 |
| | $Y_{Multi}$ | **.85** | **.80** | 20 | 5 | **.90** | 66 | 10 | **.75** | 16 | 2 | **.85** | 95 | 24 | **.79** | 35 | 14 | **1.00** | 14 | 4 | **.68** | 14 | 4 | **1.00** | 60 | 11 |
| 1E | RoB. | .10 | .20 | 20 | 5 | **.27** | 40 | 10 | **.00** | 6 | 2 | **.00** | 63 | 24 | **.19** | 29 | 14 | .01 | 7 | 4 | **.12** | 14 | 4 | **.00** | 29 | 11 |
| | Y+R. | **.11** | **.27** | 20 | 5 | .26 | 40 | 10 | **.00** | 6 | 2 | **.00** | 63 | 24 | .08 | 29 | 14 | **.17** | 7 | 4 | .11 | 14 | 4 | **.00** | 29 | 11 |
| 2A | LLM | .18 | .01 | 23 | 5 | .07 | 46 | 10 | **.00** | 8 | 2 | .33 | 96 | 24 | .00 | 18 | 14 | .09 | 12 | 4 | .11 | 23 | 4 | .84 | 32 | 11 |
| | $Y_{Mono}$ | **.65** | - | - | - | - | - | - | - | - | - | - | - | - | - | - | - | - | - | - | - | - | - | - | - | - |
| | Y+LLM | .29 | **.20** | 23 | 5 | **.20** | 46 | 10 | **.00** | 8 | 2 | .67 | 96 | 24 | .06 | 18 | 14 | **.20** | 12 | 4 | .25 | 23 | 4 | .77 | 32 | 11 |
| | $Y_{Multi}$ | .35 | **.20** | 23 | 5 | .00 | 46 | 10 | **.00** | 8 | 2 | **.73** | 96 | 24 | **.07** | 18 | 14 | .20 | 12 | 4 | **.65** | 23 | 4 | **.97** | 32 | 11 |
| 2B | LLM | .10 | **.00** | 0 | 5 | .07 | 26 | 10 | **.00** | 10 | 2 | .21 | 32 | 24 | **.00** | 6 | 14 | .27 | 7 | 4 | **.00** | 0 | 4 | .23 | 31 | 11 |
| | $Y_{Mono}$ | **.52** | - | - | - | - | - | - | - | - | - | - | - | - | - | - | - | - | - | - | - | - | - | - | - | - |
| | Y+LLM | .16 | **.00** | 0 | 5 | **.16** | 26 | 10 | **.00** | 10 | 2 | .21 | 32 | 24 | **.00** | 6 | 14 | **.50** | 7 | 4 | **.00** | 0 | 4 | .45 | 31 | 11 |
| | $Y_{Multi}$ | .20 | **.00** | 0 | 5 | .00 | 26 | 10 | **.00** | 10 | 2 | **.70** | 32 | 24 | **.00** | 6 | 14 | .33 | 7 | 4 | **.00** | 0 | 4 | **.58** | 31 | 11 |
| 2C | LLM | .01 | **.00** | 7 | 5 | .02 | 9 | 10 | **.00** | 4 | 2 | .00 | 96 | 24 | .00 | 9 | 14 | .00 | 5 | 4 | .03 | 8 | 4 | .00 | 16 | 11 |
| | $Y_{Mono}$ | **.39** | - | - | - | - | - | - | - | - | - | - | - | - | - | - | - | - | - | - | - | - | - | - | - | - |
| | Y+LLM | .06 | **.00** | 7 | 5 | .02 | 9 | 10 | **.00** | 4 | 2 | .00 | 96 | 24 | .06 | 9 | 14 | **.25** | 5 | 4 | .17 | 8 | 4 | .00 | 16 | 11 |
| | $Y_{Multi}$ | .26 | **.00** | 7 | 5 | **.20** | 9 | 10 | **.00** | 4 | 2 | **.70** | 96 | 24 | **.07** | 9 | 14 | .00 | 5 | 4 | **.50** | 8 | 4 | **.58** | 16 | 11 |
| 3A | LLM | .41 | - | 0 | 0 | .60 | 19 | 3 | - | 0 | 0 | .26 | 120 | 19 | .06 | 17 | 3 | .83 | 7 | 2 | .50 | 14 | 1 | **1.00** | 32 | 6 |
| | $Y_{Mono}$ | **.90** | - | - | - | - | - | - | - | - | - | - | - | - | - | - | - | - | - | - | - | - | - | - | - | - |
| | Y+LLM | .63 | - | 0 | 0 | **1.00** | 19 | 3 | - | 0 | 0 | **1.00** | 120 | 19 | .06 | 17 | 3 | **1.00** | 7 | 2 | **1.00** | 14 | 1 | **1.00** | 32 | 6 |
| | $Y_{Multi}$ | .58 | - | 0 | 0 | **1.00** | 19 | 3 | - | 0 | 0 | **1.00** | 120 | 19 | **.67** | 17 | 3 | **1.00** | 7 | 2 | .00 | 14 | 1 | **1.00** | 32 | 6 |
| 3B | LLM | .35 | - | 0 | 0 | .25 | 31 | 4 | **.00** | 1 | 1 | .44 | 126 | 19 | .00 | 10 | 2 | .12 | 2 | 3 | **1.00** | 36 | 5 | **1.00** | 31 | 6 |
| | $Y_{Mono}$ | **1.00** | - | - | - | - | - | - | - | - | - | - | - | - | - | - | - | - | - | - | - | - | - | - | - | - |
| | Y+LLM | .43 | - | 0 | 0 | .08 | 31 | 4 | **.00** | 1 | 1 | .95 | 126 | 19 | .00 | 10 | 2 | **.87** | 2 | 3 | .76 | 36 | 5 | .82 | 31 | 6 |
| | $Y_{Multi}$ | .54 | - | 0 | 0 | **1.00** | 31 | 4 | **.00** | 1 | 1 | **1.00** | 126 | 19 | **.50** | 10 | 2 | .00 | 2 | 3 | **1.00** | 36 | 5 | .83 | 31 | 6 |
| 4A | LLM | .52 | .61 | 96 | 20 | .61 | 363 | 57 | .54 | 60 | 18 | .32 | 1248 | 254 | .10 | 163 | 58 | .44 | 88 | 33 | **.65** | 273 | 57 | .90 | 379 | 92 |
| | $Y_{Mono}$ | **.87** | - | - | - | - | - | - | - | - | - | - | - | - | - | - | - | - | - | - | - | - | - | - | - | - |
| | Y+LLM | .57 | .45 | 96 | 20 | .57 | 363 | 57 | **.66** | 60 | 18 | .73 | 1248 | 254 | .15 | 163 | 58 | .62 | 88 | 33 | .48 | 273 | 57 | .89 | 379 | 92 |
| | $Y_{Multi}$ | .79 | **.81** | 96 | 20 | **.96** | 363 | 57 | .61 | 60 | 18 | **.90** | 1248 | 254 | **.67** | 163 | 58 | **.85** | 88 | 33 | .61 | 273 | 57 | **.92** | 379 | 92 |
| 4B | LLM | .44 | .45 | 32 | 8 | .81 | 102 | 18 | **.00** | 7 | 3 | .34 | 275 | 46 | .19 | 53 | 12 | .42 | 26 | 4 | .35 | 53 | 9 | .94 | 96 | 16 |
| | $Y_{Mono}$ | **.93** | - | - | - | - | - | - | - | - | - | - | - | - | - | - | - | - | - | - | - | - | - | - | - | - |
| | Y+LLM | .65 | .46 | 32 | 8 | .90 | 102 | 18 | **.00** | 7 | 3 | **.98** | 275 | 46 | .51 | 53 | 12 | **.50** | 26 | 4 | .88 | 53 | 9 | **1.00** | 96 | 16 |
| | $Y_{Multi}$ | .74 | **.85** | 32 | 8 | **.97** | 102 | 18 | **.00** | 7 | 3 | .96 | 275 | 46 | **.64** | 53 | 12 | **.50** | 26 | 4 | **1.00** | 53 | 9 | **1.00** | 96 | 16 |
| 4C | LLM | .31 | **.06** | 96 | 8 | .41 | 363 | 18 | .33 | 60 | 3 | .33 | 1248 | 46 | .01 | 163 | 12 | .25 | 88 | 4 | .28 | 273 | 9 | .77 | 379 | 16 |
| | $Y_{Mono}$ | **.65** | - | - | - | - | - | - | - | - | - | - | - | - | - | - | - | - | - | - | - | - | - | - | - | - |
| | Y+LLM | .39 | .00 | 96 | 8 | .49 | 363 | 18 | **.41** | 60 | 3 | .43 | 1248 | 46 | .24 | 163 | 12 | **.31** | 88 | 4 | .28 | 273 | 9 | **1.00** | 379 | 16 |
| | $Y_{Multi}$ | .42 | .00 | 96 | 8 | **.89** | 363 | 18 | .00 | 60 | 3 | **.60** | 1248 | 46 | **.37** | 163 | 12 | .29 | 88 | 4 | **.33** | 273 | 9 | .91 | 379 | 16 |

commentary, critical apparatus and footnotes. However, the textual contents of page regions such as running headers and numbers do contain straightforward meaningful information. The former always contains identical words and the latter almost only consist of Arabic numerals. This could explain why token ablation deteriorates the model's results in these two cases.

## 6. General discussion

**YOLO$_{Mono}$ and YOLO$_{Mono}$+LayoutLM.** With a single class to predict, YOLO$_{Mono}$ unsurprisingly surpasses YOLO$_{Multi}$ and displays very encouraging results. The model is above .9 in experiments 0B, 3A and 4B, generalizes better than its rivals and even reaches perfect mAP@0.5 for experiment 3B. Though these results can already be useful for other downstream tasks like pre-OCR region detection, we leverage YOLO$_{Mono}$'s predictions and use them as a basis for LayoutLM, thereby addressing the problem of rebuilding regions. With an intriguing exception in experiment 0B, this method consistently improves LayoutLM. This result is coherent with caveats that come with our evaluation system (cf. Section 4.3). Rebuilding regions from labelled sequence can indeed lead to unwanted patchwork like schemes, as small nested clusters divide regions and build new ones. However, taking the majority class among labelled tokens in an already predicted region alleviates the harshness of region-based evaluation. This methods conveys two other remarks. First, though we applied *token classification* to enable a fair comparison with baseline settings, the fact that regions are predefined allows for implementing a *sequence classification* model, which could improve the results. Indeed, it may be tough for the model to correctly label a single page number token lost in a long sequence of text. However, classifying an isolated line or page number in a pre-defined region could be an easier task. As a second remark, it is worth recalling that if this approach remains inferior to YOLO$_{Multi}$ in our case, it could prove to be more efficient with less domain-specific and noisy texts. Lifting this barrier could perhaps be achieved by the use of multilingual models such as LayoutXLM or by continuing the transformers pre-training on domain-specific data, an investigation we plan to pursue in future works.

**Inter- and intra-experiment variances** For experiments 0A, 0B, 3A and 3B, we train a single YOLO$_{Multi}$ model per commentary. Despite similar training parameters comparable amounts of data, we witness a strong variance between commentaries, with a gap of .18 between 0A and 0B. If this variance might be explained by layout particularities, we are also aware that it can be caused by the sparsity of the evaluation set. To acknowledge this limitation, we indicate the number of training and evaluation instance in Table 4. This sparsity also correlates with intra-experiment variance, i.e. differences between each region's score. As splitting at page level does not allow to precisely balance all classes, some end up being poorly distributed. Besides, footnotes are extremely rare, which can explain their poor results (mean AP=.13 over the four mentioned experiments). On the contrary, more frequent classes like commentaries and running headers yield much higher results, with a mean AP of .98 and 0.93 respectively.

## 7. Conclusions and further work

Our main contributions lie in our experiments and in the release of an annotated dataset. The key takeaways of this research are listed below:

- We show that an object detection model such as YOLO succeeds in classifying semantic regions of text-heavy documents even if they feature little obvious graphical differences.

- Hybrid models like LayoutLM may be of help to researchers working with clean and generic English data. However, in a highly noisy, multi-lingual, domain-specific and historic setting, they tend to make little use of the textual channel and mainly base their decision on coordinates and images.
- With 8 classes in total, we show that annotating 15 to 20 pages of ground truth data already yields satisfactory results. Doubling this amount amends results by .05% in average.

Furthermore, because historical classical commentaries and critical editions have a significant similarity in terms of layout, our approach establishes the groundwork for developing a robust, generic model for page layout analysis of these publications in the near future. Such a model, in combination with existing open source tools for annotation that can be chained into a seamless pipeline (e.g. eScriptorium for annotation and Kraken for OCR), has the potential to open up new perspectives to researchers for exploiting openly available digitized editions and commentaries. Similarly, such a model could be useful to projects aimed at the creation of large-scale corpora of marked up texts such the Free First Thousands Years of Greek (FF1KG) project [21]. In this project, dozens of summer interns over the years have manually annotated the layout of digitized critical editions, carrying out a tedious task that in the near future will be possible to semi-automate.

As for future works, we think two strands of possible improvement are worth investigating. First, and as mentioned in Section 5, we would like to explore the effect of adding minimal in-domain data when fine-tuning a generic model. Based on our experiments (notably 1A), our hypothesis is that providing only a few pages should improve the results on unseen commentaries. Secondly, we think our hybrid would yield better results if they could use more meaningful representation of the text. It could therefore be worth testing with a multilingual model such as LayoutXLM or with a domain-specific language modelling pre-training.

## Acknowledgments

This research has been supported by the Swiss National Science Foundation under an Ambizione grant PZ00P1_186033. We thank Carla Amaya who helped in the ground truth annotation process.

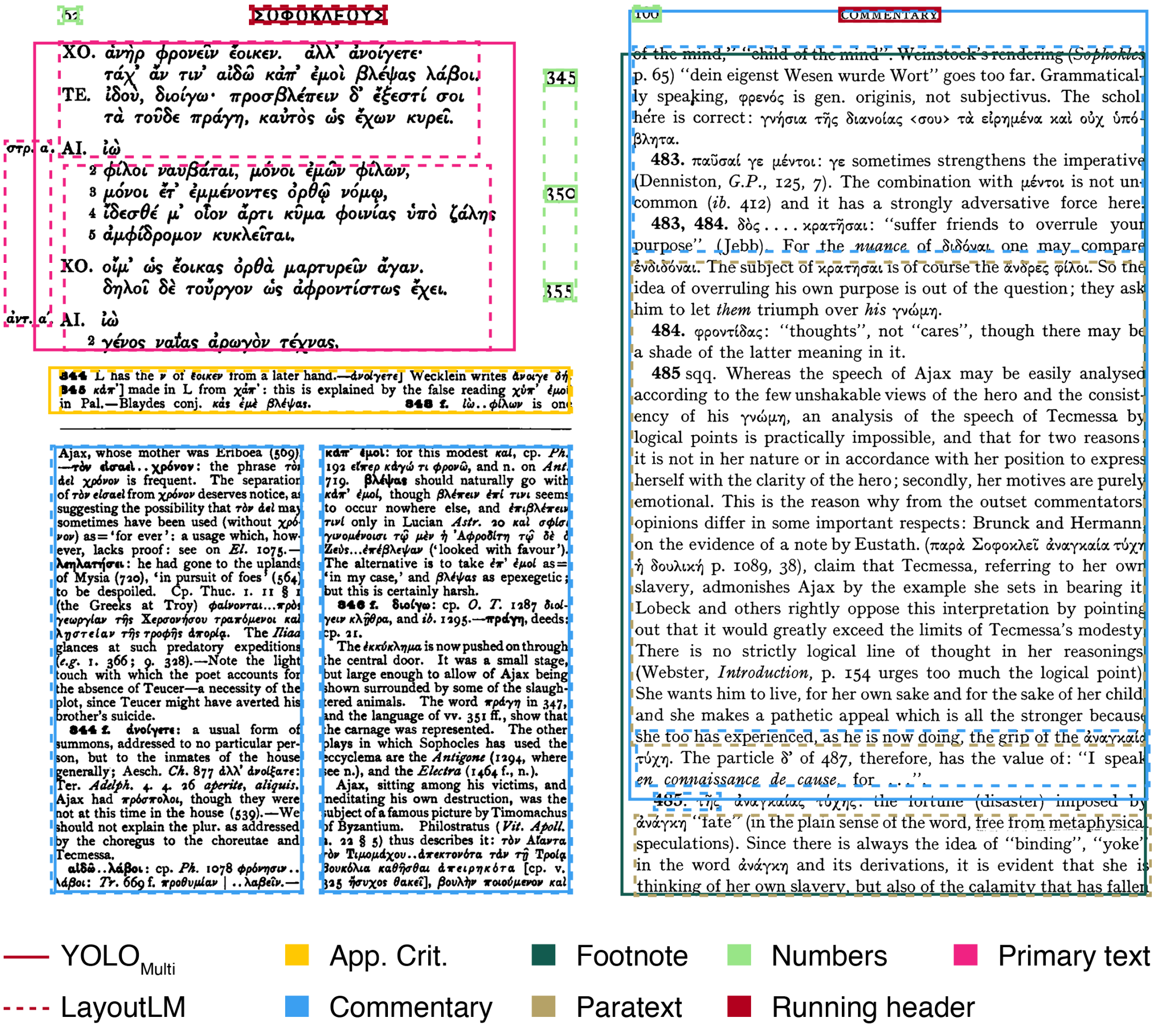

**Figure 4:** Examples of pages by Jebb (left) and Kamerbeek (right) with the predictions of LayoutLM's and YOLO$_{Multi}$'s best checkpoints from experiment 4A.